\documentclass{llncs}

\usepackage{algorithm}
\usepackage{algpseudocode}
\usepackage{graphicx}

\begin{document}

\title{Some Extensions of the All Pairs Bottleneck Paths Problem\thanks{This research was supported by the EU/NZ Joint Project, Optimization and its Applications in Learning and Industry (OptALI).}}

\author{Tong-Wook Shinn \and Tadao Takaoka}
\institute{Department of Computer Science and Software Engineering\\
University of Canterbury\\
Christchurch, New Zealand}

\maketitle

\begin{abstract}
We extend the well known bottleneck paths problem in two directions for directed unweighted (unit edge cost) graphs with positive real edge capacities. Firstly we narrow the problem domain and compute the bottleneck of the entire network in $O(n^{\omega}\log{n})$ time, where $O(n^{\omega})$ is the time taken to multiply two $n$-by-$n$ matrices over ring. Secondly we enlarge the domain and compute the shortest paths for all possible flow amounts. We present a combinatorial algorithm to solve the Single Source Shortest Paths for All Flows (SSSP-AF) problem in $O(mn)$ worst case time, followed by an algorithm to solve the All Pairs Shortest Paths for All Flows (APSP-AF) problem in $O(\sqrt{d}n^{(\omega+9)/4})$ time, where $d$ is the number of distinct edge capacities. We also discuss real life applications for these new problems.
\end{abstract}

\section{Introduction}
The bottleneck (capacity) of a path is the minimum capacity of all edges on the path. Thus the bottleneck is the maximum flow that can be pushed through this path. The bottleneck (capacity) of a pair of vertices $(i,j)$ is the maximum of all bottleneck values of all paths from $i$ to $j$. The bottleneck paths problems are important in various areas, such as logistics and computer networks. In this paper we consider two extensions to this well known problem on directed unweighted graphs with positive real edge capacities.

The bottleneck of the (entire) network is the minimum bottleneck out of all bottlenecks for all pairs $(i,j)$. In this paper we introduce a simple algorithm based on binary search to show that we can compute the bottleneck of the network faster than computing the all pairs bottleneck paths. The method is based on the transitive closure of a Boolean matrix and the time complexity of the algorithm is $O(n^{\omega}\log{n})$ where $\omega=2.373$ \cite{Williams}. This algorithm is simple but effective, and provides a good starting point for this paper.

Consider the shortest path from vertex $s$ to vertex $t$ that can push a flow of amount up to $b$. If the flow demand from $s$ to $t$ is less than $b$, however, there may be a shorter route, which is useful if one wishes to minimize the distance for a given amount of flow. Thus we compute the shortest path for each possible bottleneck value. We call this problem Shortest Paths for All Flows (SP-AF). We present a non-trivial $O(mn)$ algorithm to solve the Single Source Shortest Paths for All Flows (SSSP-AF) problem, that is, computing the shortest path for all flows from one source vertex to all other vertices in the graph.

Naturally, we move onto the All Pairs Shortest Paths for All Flows (APSP-AF) problem, where we compute the shortest distances for all flows for all pairs of vertices in the graph. Note that this new problem is different from the All Pairs Bottleneck Shortest Paths (APBSP) problem \cite{VWY}, which is to compute the bottlenecks for all pairs shortest paths. Applying our algorithm for SSSP-AF $n$ times gives us $O(mn^{2})$. If the graph is dense, however, $m = O(n^{2})$, and the time complexity becomes $O(n^{4})$. We can utilize faster matrix multiplication over ring to achieve a sub-quartic time bound for dense graphs. We present an algorithm that runs in $O(\sqrt{m}n^{(\omega+9)/4})$ time. If the edge capacities are integers bounded by $c$, the time complexity of our algorithm becomes $O(\sqrt{min\{c,m\}}n^{(\omega+9)/4})$, and if we introduce the parameter $d$ as the number of distinct edge capacities, we have $O(\sqrt{d}n^{(\omega+9)/4})$.

The algorithms are presented in the order of increasing complexity. Section \ref{sec:bpen} details the algorithm for solving the bottleneck of the network. In Section \ref{sec:ssspaf} and Section \ref{sec:apspaf} we present the algorithms for solving SSSP-AF and APSP-AF, respectively. In Section \ref{sec:distinct} we further analyze our new algorithm for APSP-AF. Finally we describe some real life applications of the SP-AF problems in Section \ref{sec:real} before concluding the paper.

\section{Preliminaries}
\label{sec:prelim}
Let $G=\{V,E\}$ be a directed unweighted graph with edge capacities of positive real numbers. Let $n = |V|$ and $m = |E|$. Vertices (or nodes) are given by integers such that $\{1,2,3,...,n\} \in V$. Let $e(i,j) \in E$ denote the edge from vertex $i$ to vertex $j$. Let $cap(i,j)$ denote the capacity of the edge $e(i,j)$.

We call $C=\{c_{ij}\}$, where $c_{ij}$ represents a capacity from $i$ to $j$, a capacity matrix. Let $C^{\ell} = c^{\ell}_{ij}$ be the maximum bottleneck for all paths of lengths up to $\ell$ from $i$ to $j$. Clearly $c^{1}_{ij} = cap(i,j)$ if $e(i,j) \in E$, and 0 otherwise. Let $c^{*}_{ij}$ be the maximum bottleneck for all paths from $i$ to $j$. We call $C^*=\{c^{*}_{ij}\}$ the closure of $C$ and also refer to it as the bottleneck matrix. The problem of computing $C^*$ is formally known as the All Pairs Bottleneck Paths (APBP) problem. For graphs with unit edge costs, the APBP problem is well studied in \cite{VWY} and \cite{DP}. The complexities of algorithms given by the two papers are $O(n^{2+\omega/3}) = O(n^{2.791})$ and $O(n^{(\omega+3)/2}) = O(n^{2.687})$, respectively.

Let $Q=A \star B$ denote the $(max,min)$-product of capacity matrices $A$ and $B$, where $Q=\{q_{ij}\}$ is given by:

$$
q_{ij} = \max_{1 \leq k \leq n}\{\min\{a_{ik}, b_{kj}\}\}
$$

\noindent
Note that if all elements in $A$ and $B$ are either 0 or 1, this becomes Boolean matrix multiplication. If we interpret ``max'' as addition and ``min'' as multiplication, the set of non-negative numbers forms a closed semi-ring. Similarly, the set of matrices where the product is defined as the $(max,min)$-product and the sum is defined as a component-wise ``max'' operation also forms a closed semi-ring. Then the bottleneck matrix is given by the closure of the capacity matrix, where the closure of matrix $A$ is defined by:

$$
A^* = I + A + A^2 + A^3 + ...
$$

\noindent
and $I$ is the identity matrix with diagonal elements of $\infty$ and non-diagonal elements of 0. Although $A^*$ is defined by an infinite series we can stop at $n-1$. The computational complexity of computing $A^*$ is asymptotically equal to that of the matrix product in the more general setting of closed semi-ring \cite{Aho}.

Similarly to the capacity matrix, we can define the distance matrix, where each element represents the distance from $i$ to $j$. The problem of computing the closure of the distance matrix is formally known as the All Pairs Shortest Paths (APSP) problem. Zwick achieved $O(n^{2.575})$ time for solving APSP on directed graphs with unit edge costs \cite{Zwick}, which has recently been improved to $O(n^{2.53})$ thanks to Le Gall's new algorithm for rectangular matrix multiplication \cite{Gall}.

Let $Q=A \ast B$ denote the $(min,+)$-product, or the distance product, of distance matrices $A$ and $B$, where $Q=\{q_{ij}\}$ is given by:

$$
q_{ij} = \min_{1 \leq k \leq n}\{a_{ik} + b_{kj}\}
$$

\section{The bottleneck problem of the entire network}
\label{sec:bpen}
Let $bottleneck$ be the bottleneck value of the entire network. See Example \ref{eg:bottleneck} for an illustration. Let the capacity matrix C be defined by $c_{ij} = cap(i,j)$. One straightforward method to compute $bottleneck$ would be to compute $C^{*} = \{c^{*}_{ij}\}$ for all pairs $(i,j)$ and find the minimum among them. We can solve the problem more efficiently by a simple but effective binary search as shown in Algorithm \ref{alg:bpen}.

We begin by assuming that the edge capacities are integers bounded by $c$. Let the threshold value $t$ be initialized to $c/2$. Let $B = \{b_{ij}\}$ be a Boolean matrix such that $b_{ij}=1$ if $cap(i,j) \ge t$, and 0 otherwise. Let us compute the transitive closure, $B^*$, of $B$. Then, from the equation:

$$
b^{*}_{ij}=\Sigma \{b_{ik_{1}}b_{k_{1}k_{2}}...b_{k_{r}j}\mbox{ } | \mbox{ all possible paths } e(i,k_{1}), e(k_{1},k_{2}), ..., e(k_{r},j)\}
$$

\noindent
we observe that $b^{*}_{ij}=1$ if and only if $b_{ik_{1}}=1$, $b_{k_{1}k_{2}}=1$, ..., $b_{k_{r}j}=1$ for some path. From this we derive that $bottleneck \ge t$ \emph{iff} $b^{*}_{ij} > 0$ for all pairs $(i,j)$. We repeatedly halve the possible range $[\alpha, \beta]$ for $bottleneck$ by adjusting the threshold, $t$, through binary search.

\begin{algorithm}
\caption{Solve bottleneck problem of the entire network}
\label{alg:bpen}
\begin{algorithmic}[1]
\algnotext{EndFor}
\algnotext{EndIf}
\algnotext{EndWhile}
\State{$\alpha \leftarrow 0$}
\State{$\beta \leftarrow c$}
\While{$\beta - \alpha > 0$}
	\State{$t \leftarrow (\alpha + \beta)/2$}
	\For{$i \leftarrow 1$ to $n$, $j \leftarrow 1$ to $n$}
		\If{$c_{ij} \ge t$}
			\State{$b_{ij} \leftarrow 1$}
		\Else
			\State{$b_{ij} \leftarrow 0$}
		\EndIf
	\EndFor
	\State{Compute $B^*$}
	\If{$b^{*}_{ij} > 0$ for all $i,j$}
		\State{$\alpha \leftarrow t$}
	\Else
		\State{$\beta \leftarrow t$}
	\EndIf
\EndWhile
\State{$bottleneck \leftarrow \alpha$}
\end{algorithmic}
\end{algorithm}

Obviously the iteration over the while loop is performed $O(\log{c})$ times. Thus the total time becomes $O(B(n)\log{c})$, where $B(n)$ is the time for multiplying two $n$-by-$n$ Boolean matrices. If $c$ is large, say $O(2^n)$, the algorithm is not very efficient, taking $O(n)$ halvings of the possible ranges of $bottleneck$. In this case, we sort edges in ascending order. Since there are at most $m$ possible values of capacities, doing binary search over the sorted edges gives us $O(n^{\omega}\log{m}) = O(n^{\omega}\log{n})$. Obviously this method also works for edge capacities of real numbers. We note that the actual bottleneck path can be obtained using the witness technique in \cite{AGM} with an extra polylog factor.

\begin{figure}
\centering
\includegraphics[scale=0.35]{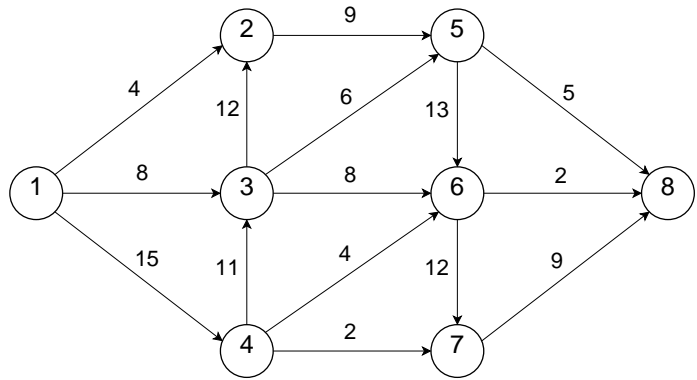}
\caption{An example of directed unweighted graph with edge capacities.}
\label{fig:graph}
\end{figure}

\begin{example}
\label{eg:bottleneck}
The $bottleneck$ of the graph in Figure \ref{fig:graph} is 9, which is the capacity of edges $e(2,5)$ and $e(7,8)$. This example illustrates that the $bottleneck$ of an entire network, even for simple graphs, may not be immediately obvious.
\end{example}

\section{Single source shortest paths for all flows problem}
\label{sec:ssspaf}
From a source vertex $s$ to all other vertices $v \in V$, we want to find the shortest paths for each flow value. The shortest path from $s$ to $v$ for a given flow value $f$ allows us to push flows up to $f$ as quickly as possible. For some $f' < f$, however, there may be a shorter path. Thus if we find the shortest path for all possible flows, we can respond to queries of flow demands from $s$ to $v$ with the quickest paths that can accommodate the flows. We observe that there can be up to $m$ different values of $f$, which we refer to as the maximal flows from $s$ to $v$.

A straightforward method of solving SSSP-AF is solving SSSP for each maximal flow $f$, that is, we repeatedly solve SSSP using only $e(u,v) \in E$ such that $cap(u,v) \geq f$, for all $f$. SSSP can be solved by a simple breadth-first-search (BFS) on graphs with unit edge costs, hence this approach takes $O(m^{2})$ time. Each BFS will result in a shortest path spanning tree (SPT) with $s$ as the root. Explicit paths can be retrieved by traversing up the SPTs.

One may be led to think that SSSP-AF can be solved with a simple decremental algorithm, that is, repeatedly removing edges in decreasing order of capacity, and checking for connectivity of vertices. This method, however, gives incorrect results because edges with larger capacities may later be required to provide shorter paths for smaller flows. The SP-AF problem requires solving the shortest paths problem and the bottleneck paths problem at the same time. This is not a trivial matter, as operations required to solve the two problems generally take us in opposite directions; maximizing bottlenecks comes at the cost of increased distances and minimizing distances comes at the expense of decreased bottlenecks. We have achieved $O(mn)$ worst case time for solving SSSP-AF by fully exploiting the fact that all edges have unit costs. The resulting algorithm was surprisingly simple, as shown in Algorithm \ref{alg:ssspaf}.

\begin{algorithm}
\caption{Solve single source shortest paths for all flows problem}
\label{alg:ssspaf}
\begin{algorithmic}[1]
\algnotext{EndFor}
\algnotext{EndIf}
\algnotext{EndWhile}
\For{$i \leftarrow 1$ to $n$}
	\State{$B[i] \leftarrow 0, L[i] \leftarrow 0$}
\EndFor
\State{$B[s] \leftarrow \infty, T \leftarrow s$} /* $T$ is for $SPT$, initially only root $s$ */
\ForAll{maximal flows $f$ in increasing order}
	\ForAll{$v \in V$ such that $B[v] < f$}
		\If {$v$ is in $T$}
			\State{Cut $v$ from $T$}
		\EndIf
		\State{$L[v] \leftarrow L[v] + 1$}
		\State{Push $v$ to $Q[L[v]]$} /* $v$ to be processed later */
	\EndFor
	\For{$\ell \leftarrow 1$ to $n-1$}
		\While{$Q[\ell]$ is not empty}
			\State{Pop $v$ from $Q[\ell]$}
			\ForAll{$e(u,v) \in E$}
				\If{$L[u] = L[v] - 1$}
					\If{$\Call{min}{cap(u,v), B[u]} > B[v]$}
						\State{$B[v] \leftarrow \Call{min}{cap(u,v), B[u]}$} /* $B[v]$ increased */
						\State{Add $v$ to $T$ with $u$ as the parent}
					\EndIf
				\EndIf
			\EndFor
			\If{$v$ is not in $T$}
				\State{$L[v] \leftarrow L[v] + 1$}
				\State{Push $v$ to $Q[L[v]]$} /* $v$ to be processed later */
			\EndIf
		\EndWhile
	\EndFor
\EndFor
\end{algorithmic}
\end{algorithm}

Let $B[v]$ be the bottleneck of a path from $s$ to vertex $v$, $L[v]$ be the possible length of the path from $s$ to $v$, and let $T$ represent the SPT. $T$ is a kind of persistent data structure, that is, we do not compute $T$ from scratch for each maximal flow. Let $Q[i]$ be a set of vertices that may be added to $T$ at distance $i$, such that $1 \leq i \leq n - 1$, i.e. one set for each possible path length from $s$. We iterate through each maximal flow $f$ in increasing order. At each iteration, all $v \in V$ such that $B[v] < f$ is cut from $T$ and added to $Q[L[v]+1]$. The key observation here is that when $v$ gets cut from $T$, it is only possible for $v$ to be re-added to $T$ at a greater distance from $s$ than the previous distance. For all vertices that have been cut, we attempt to add each back to $T$ at the minimum possible path length from $s$ for current $f$ by emptying $Q$ from $Q[1]$ to $Q[n-1]$. If there are many potential parent nodes at a given path length, we choose the parent that gives us the maximum bottleneck. $T$ is organized as a linked structure with detailed operations being omitted. At each iteration we are effectively solving SSSP for the given value of $f$ by performing incremental updates to the SPT. Therefore at the end of each iteration $L[*]$ contains the shortest distances for all destination vertices for maximal flow $f$, and the path can be retrieved by traversing up the SPT.

We perform lifetime analysis of vertices in the data structure $Q[*]$ for the worst case time complexity of Algorithm \ref{alg:ssspaf}. Each vertex $v$ can be cut from $T$ and be re-added to $T$ $O(n)$ times, once per each possible path length from $s$. Cutting/adding $v$ from/to $T$ takes $O(1)$ time, achieved by setting the parent of $v$ to either \emph{null} or $u$, respectively. Therefore the total time complexity of all operations involving $T$ is $O(n^2)$. Before each vertex $v$ is added to $T$ all incoming edges $e(u,v)$ are inspected. This results in $O(m)$ edges being inspected in total for the entire duration of the algorithm for each possible path length from $s$. Since there are $O(n)$ possible path lengths, the total time taken for edge inspection is $O(mn)$. Even though we iterate up to $O(m)$ times, we are bounded by the fact that each vertex can only be observed at each possible path length exactly once. Therefore the total time complexity of the algorithm is $O(mn)$. Obviously APSP-AF can be
solved in $O(mn^{2})$ by running this algorithm $n$ times.

\begin{figure}
\centering
\includegraphics[scale=0.33]{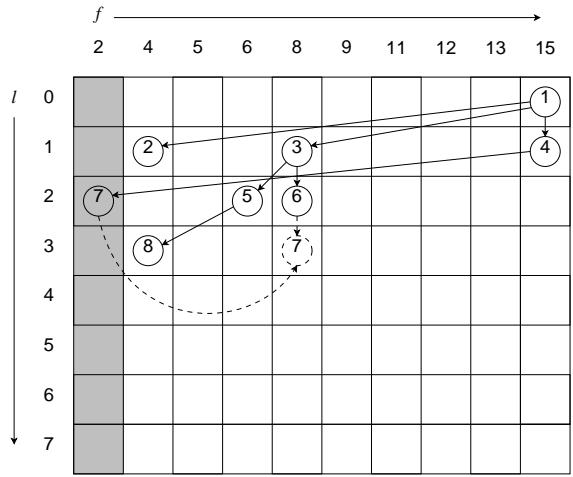}
\caption{Changes to the SPT at iteration $f=4$.}
\label{fig:spt}
\end{figure}

\begin{example}
Figure \ref{fig:spt} shows Algorithm \ref{alg:ssspaf} being applied on the graph in Figure \ref{fig:graph} with $s=1$. Initially $T$ is created with all edges. At iteration $f=4$, $e(4,7)$ is cut, causing vertex 7 to be reattached to $T$ under vertex 6. $L[7]$ is increased from 2 to 3 and $B[7]$ is increased from 2 to 8. Vertices in $T$ will never reoccupy the shaded region, which will grow larger to the right as $f$ increases.
\end{example}

\section{All pairs shortest paths for all flows problem}
\label{sec:apspaf}
For each pair of vertices $(i,j)$ for each maximal flow, we want to compute the shortest path. Thus our aim here is to obtain tuples of pairs $(\ell,f)$ for all $(i,j)$, where $f$ is the maximum flow that can be pushed through a shortest path whose length is $\ell < n$. We can assume that the values of $\ell$ are all distinct.

Let $C$ be the capacity matrix and let $D^{f}$ be the approximate distance matrix for paths that can accommodate flows up to $f$. A more detailed description of $D^{f}$ follows shortly in the main description of our algorithm. Let $T$ be a matrix such that $t_{ij}$ is a tuple of pairs $(\ell,f)$ as described above. Let both $(\ell,f)$ and $(\ell',f')$ be in $t_{ij}$ such that $\ell < \ell'$. We keep $(\ell',f')$ \emph{iff} $f < f'$ i.e. a longer path is only relevant if it can accommodate a greater flow. Each $t_{ij}$ has at most $n-1$ pairs of $(\ell, f)$. We assume the pairs are sorted in ascending order of $\ell$. We make an interesting observation here that the set of first pairs for all $t_{ij}$ is the solution to APBSP, and the set of last pairs for $t_{ij}$ is the solution to APBP. For APSP-AF, all pairs $(\ell,f)$ for all $t_{ij}$ are computed.

\begin{example}
Solving APSP-AF on the graph given in Figure \ref{fig:graph} results in a tuple of four pairs of $(\ell, f)$ from vertex 4 to vertex 7, that is, $t_{47} = ((1,2),(2,4),(3,8),(5,9))$.
\end{example}

\begin{algorithm}
\caption{Solve all pairs shortest paths for all flows problem}
\label{alg:apspaf}
\begin{algorithmic}[1]
\algnotext{EndFor}
\algnotext{EndIf}
\algnotext{EndWhile}
\Statex{/* initialization for acceleration phase */}
\State{$C^{0} \leftarrow I$}
\For{$i \leftarrow 1$ to $n$; $j \leftarrow 1$ to $n$}
	\State{$t_{ij} \leftarrow \phi$} /* $\phi$ is empty */
\EndFor

\Statex{/* acceleration phase */}
\For{$\ell \leftarrow 1$ to $r$}
	\State{$C^{\ell} \leftarrow C^{\ell-1} \star C$}
	\For{$i \leftarrow 1$ to $n$; $j \leftarrow 1$ to $n$ such that $i \neq j$}
		\State{$f \leftarrow c^{\ell}_{ij}$}
		\If{$f > c^{\ell-1}_{ij}$}
			\State{$t_{ij} \leftarrow t_{ij} || (\ell,f)$} /* append $(\ell,f)$ to $t_{ij}$ */
		\EndIf
	\EndFor
\EndFor
\Statex{/* initialization for cruising phase */}
\For{$i \leftarrow 1$ to $n$; $j \leftarrow 1$ to $n$ such that $i \neq j$}
	\ForAll{$x$ in $T[i,j]$}
		\If{$x \neq \phi$}
			\State{let $x=(\ell,f)$}
			\State{$d^{f}_{ij} \leftarrow \ell$}
		\Else
			\State{$d^{f}_{ij}\leftarrow \infty$}
		\EndIf
	\EndFor
\EndFor
\Statex{/* cruising phase */}
\State{$\ell \leftarrow r$}
\While{$\ell < n$}
	\State{$\ell_{1} \leftarrow \lceil \ell * 3/2 \rceil$}
	\ForAll{maximal flow $f$}\label{line:iterate}
		\For{$i \leftarrow 1$ to $n$}
			\State{Scan $i^{th}$ row of $D^{f}$ with $j$ and find the smallest set of equal $d^{f}_{ij}$ }
			\State{\mbox{ } \mbox{ } such that $\lceil \ell/2 \rceil \leq d^{f}_{ij} \leq \ell$ and let the set of corresponding $j$ be $S_{i}$}
		\EndFor
		\For{$i \leftarrow 1$ to $n$; $j \leftarrow 1$ to $n$ such that $i \neq j$}
			\State{$m_{ij} \leftarrow \min_{k \in S_{i}}\{d^{f}_{ik} + d^{f}_{kj}\}$}
			\If{$m_{ij} \leq \ell_{1}$}
				\State{$d^{f}_{ij} \leftarrow m_{ij}$}
			\EndIf
		\EndFor
	\EndFor
	\State{$\ell \leftarrow \ell_{1}$}
\EndWhile
\Statex{/* finalization */}
\For{$i \leftarrow 1$ to $n$; $j \leftarrow 1$ to $n$ such that $i \neq j$}
\ForAll{maximal flow $f$ in increasing order}
\State{$\ell \leftarrow d^{f}_{ij}$}
 \State{Let the last pair of $t_{ij}$ be $x=(\ell', f')$}
    \If{$x = \phi$ or $(f > f'$ and $\ell < \infty)$}
		\If{$\ell = \ell'$}
			\State{Replace $x$ with $(\ell,f)$}
		\Else
			\State{$t_{ij} \leftarrow t_{ij} || (\ell, f)$} /* append $(\ell, f)$ to $t_{ij}$ */
		\EndIf
    \EndIf
   \EndFor
\EndFor
\end{algorithmic}
\end{algorithm}

Algorithm \ref{alg:apspaf} is largely based on the method given by Alon, Galil and Margalit in \cite{AGM}, which is commonly used to solve various all pairs path problems \cite{GM,Takaoka,Zwick,VWY}. This method has been reviewed in \cite{Takaoka} and we use the same set of terminologies as the review. The algorithm consists of two phases; the \emph{acceleration} phase and the \emph{cruising} phase. Simply speaking, we run the algorithm by Alon \emph{et al.} for all $f$ in parallel with a modified acceleration phase.

We compute the $(max,min)$-products in the acceleration phase, multiplying the capacity matrix $C$ one by one. The $\ell^{th}$ iteration of the acceleration phase, therefore, finds the maximum bottleneck for all paths of lengths up to $\ell$. 

After the acceleration phase we initialize distance matrices $D^{f}$ from $T$, one matrix for each maximal flow $f$, in preparation for the cruising phase. At this stage, $d^{f}_{ij}$ is the length of the shortest path that can push flow $f$, if the path length is $r$ or less. In the cruising phase, we perform repeated squaring on the distance matrices with the help of the bridging set $S_{i}$. At the end of the cruising phase we thus have the shortest paths for all flows for all $(i,j)$. Retrieving tuples of $(\ell,f)$ from the resulting matrix $D^{f}$ is a reverse process of the initialization for the cruising phase.

Now we analyze the worst case time complexity of Algorithm \ref{alg:apspaf}. For the acceleration phase we use the the current best known algorithm given by Duan and Pettie \cite{DP} to compute the $(max,min)$-product in each iteration, giving us $O(rn^{(3+\omega{})/2})$. The time complexity for the cruising phase is $O(mn^3/r)$. This is because $|S_{i}|$ is $O(n/r)$ as proven in \cite{AGM}, and no logarithmic factor is required for repeated squaring because the path length $\ell$ increases by a factor of $\frac{3}{2}$ in each iteration and hence the first squaring dominates the complexity. The time complexity for the initialization for the cruising phase and the finalization is $O(mn^2)$, which is absorbed by $O(mn^3/r)$ since $n/r > 1$. We balance the time complexities of the two phases by setting $r = \sqrt{m}n^{(3-\omega)/4}$, which gives us the total time complexity of $O(\sqrt{m}n^{{(\omega+9)/4}})$. If capacities are integers bounded by $c$, we only have to iterate $c$ times in line \ref{line:iterate}, giving us $O(\sqrt{min\{c, m\}}n^{(\omega+9)/4})$.

As noted earlier there can be up to $n-1$ $(\ell,f)$ pairs for each vertex pair $(i,j)$. Since the lengths of each path can be $O(n)$, explicitly listing all paths could take $O(n^4)$ time. We get around this by extending the pair $(\ell,f)$ to the triplet $(\ell,f,u)$, where $u$ is the successor node. In the acceleration phase witnesses can be retrieved with an extra polylog factor \cite{DP}, and the successor nodes can be computed from the witnesses at each iteration in $O(n^2)$ time \cite{Zwick}. In the cruising phase retrieving $u$ is a trivial exercise since ordinary matrix multiplication is performed. We can generate explicit paths in time linear to the path length by using $\ell$ as the index for looking up subsequent successor nodes. That is, we can still retrieve each successor node in $O(1)$ time even with $O(n)$ triplets $(\ell,f,u)$ for all pairs $(i,j)$ because we know that the path length decrements by 1 as we step through each successor node.

\section{Distinct edge capacities -- parameter $d$}
\label{sec:distinct}
So far we have been assuming that the number of distinct edge capacities, $d$, is bounded by the number of edges, $m$, and used only $m$ and $n$ as complexity parameters. Hence the worst case time complexity of Algorithm \ref{alg:apspaf} was given to be $O(\sqrt{m}n^{{(\omega+9)/4}})$. If we compare this with $O(mn^{2})$ given by Algorithm \ref{alg:ssspaf}, and $O(n^{(5+\omega)/2})$ given by simply staying in the acceleration phase of Algorithm \ref{alg:apspaf} until $r=n$, then we observe that for dense graphs Algorithm \ref{alg:apspaf} is faster than $O(mn^{2})$, and for sparse graphs Algorithm \ref{alg:apspaf} is faster than $O(n^{(5+\omega)/2})$.

As we will discuss further in Section \ref{sec:real}, $d$ may not be related to $m$, especially for dense graphs where $m = O(n^{2})$. Therefore we incorporate $d$ directly into the time complexity of Algorithm \ref{alg:apspaf} to give $O(\sqrt{d}n^{{(\omega+9)/4}})$ and the merit of Algorithm \ref{alg:apspaf} emerges with dense graphs having relatively small number of maximal flows. Another straightforward method of solving APSP-AF on graphs with $d$ distinct edge capacities is to compute the $(min,+)$-closure $d$ times. Using Zwick's algorithm in \cite{Zwick}, the time complexity of this method is $O(dn^{2.53})$. Clearly, for most values of $d$, Algorithm \ref{alg:apspaf} is faster.

It is well known that algorithms that utilize faster matrix multiplication over ring is not practical to be used on modern day computers \cite{Robinson} and are mostly of theoretical importance. We therefore highlight that Algorithm \ref{alg:apspaf} can be easily turned into a practical algorithm by computing $(max,min)$-product in the acceleration phase using the naive approach. This results in the time complexity of $O(\sqrt{d}n^{3})$, which is still very much useful for dense graphs alongside our combinatorial algorithm of $O(mn^{2})$ that is better suited for sparse graphs.

\section{Real life applications of APSP-AF}
\label{sec:real}
Computer networks can be accurately modeled by unweighted directed graphs with edge capacities, by representing each hop (e.g. router) as a vertex, each network link as an edge, and the bandwidth of each link as edge capacities. However, routing protocols that are commonly used today are based on less accurate models. For example, the Routing Information Protocol (RIP) computes routes based solely on the hop counts, while the Open Shortest Path First (OSPF) protocol, by default, computes routes based solely on the bandwidths.

In today's computer networks each router is autonomous, and therefore each router computes SSSP. RIP is often implemented with Bellman-Ford algorithm \cite{Ford,Bellman} and OSPF is often implemented with Dijkstra's algorithm \cite{Dijkstra}. We present SSSP-AF as a better solution that uses both the hop count and the bandwidth at the same time. Advanced routers are able to gather information such as the current flow amount from one IP subnet to another. With SSSP-AF, a router can make a better routing decision for a given flow based on the flow amount by choosing a route that minimizes the latency without causing congestion.

Furthermore, we introduce APSP-AF as a potential routing algorithm for Software Defined Networking (SDN) \cite{SDN}. SDN is a new paradigm in computer networking where routers are no longer autonomous and the whole network can be controlled at a centralized location. The central controller has in-depth knowledge of the network and as a result SDN can benefit from more sophisticated routing algorithms. By solving APSP-AF for the whole network, the fastest routes can be determined for all flow requirements for all sources and destinations. As noted in Section \ref{sec:distinct}, Algorithm \ref{alg:apspaf} can be easily turned into a practical $O(\sqrt{d}n^3)$ algorithm. This is very much relevant in real life computer networks where distinct bandwidth values are defined (e.g. 100Mbps, 1Gbps).

\section{Concluding remarks}
\label{sec:conclude}
We have extended the well known bottleneck paths problems, introducing new problems that have real life applications. We provided non-trivial algorithms to solve the problems more efficiently than straightforward methods.

This paper only considered directed unweighted graphs. In enterprise computer networks, most links are bi-directional, meaning undirected graphs are adequate for modeling those networks. Also for computer networks involving low latency switches and long cables with repeaters, introducing edge costs may enable more accurate modeling of the networks. Hence solving the SP-AF problem on other types of graphs would not only be a natural extension to this paper, but also have relevance in real life.

Trivial lower bounds of $\Omega(n^2)$ and $\Omega(n^3)$ exist for SSSP-AF and APSP-AF, respectively. Most current researches in all pairs paths problems focus on breaking the cubic barrier of $O(n^3)$ to get closer to the trivial lower bound of $\Omega(n^2)$. With the APSP-AF problem we have effectively shifted the focus in time complexities from ``cubic-to-quadratic'' to ``quartic-to-cubic''. We have thus opened up a new area of research, where we hope many new contributions would occur in the future.

\end{document}